# STEAM RECOMMENDATION SYSTEM

*Using implicit and explicit indicators to generate game recommendations for users on Steam.*

## Group 1

Samin Batra
(A0262665X)
samin.batra@u.nus.edu

Xinyao Wang
(A0262715E)
e0997938@u.nus.edu

Yurou Sun
(A0231959U)
e0709214@u.nus.edu

Varun Sharma
(A0262674X)
sharmavarun.s@u.nus.edu

Yinyu Wang
(A0262769M)
e0997992@u.nus.edu

## ABSTRACT

This project aims to leverage the interactions between users and items on the Steam community to build a game recommendation system that makes personalized suggestions to players in order to boost Steam's revenue as well as improve users' gaming experience. The whole project is built on Apache Spark dealing with big data. The final output of the project is a recommendation system that gives a list of the top 5 items that the users will possibly like.

## 1  Introduction

The world of video games has changed considerably over the recent years. Steam, as one of the biggest online video distribution platforms, has well reflected this trend. According to Newzoo[1], in 2020 Steam's main market, PC games, accounted for 21% of the global game market revenue, reaching US$33.9 billion, a year-on-year increase of 6.7%. From 2016 to 2018, the average annual compound growth rate of Steam's global registered users was 54%. With the vigorous momentum in players, a large number of explosive games are pouring into steam as well. However, the fact of having such a variety of products and so many users, makes it difficult to predict whether a particular new game will be purchased or will be endorsed by a certain user. Also, according to Steam registries of 2014, about 37% of games purchased have never been played by the users who bought them. This context creates the urgent demand for building a game recommender system that is able to make relevant personalized suggestions to players, boosting Steam's revenue as well as improving users' gaming experience.

To achieve this goal, we implemented a state-of-the-art algorithm based on Collaborative Filtering (CF) that uses the Alternating Least Squares (ALS) for making recommendations. Utilizing the implicit feedback is an important component in ALS which fits well to the characteristics of our dataset where implicit indicators for users' attitudes towards games such as reviews and playing time, as well as an explicit indicator, about whether a user recommends a game or not, are accessible.

## 2  Data and Methods

### 2.1  The Datasets

To implement the game recommender system for Steam and evaluate its prediction accuracy, we used two recent datasets on user-items and user-item-reviews on https://cseweb.ucsd.edu/~jmcauley/datasets.html#steam_data shared by Julian McAuley. The users-items dataset covers the game purchase history of Australian users on Steam. Specifically, this dataset includes users' portraits such as user id, the number of games purchased, playing time and item information. On the Steam platform, users can post reviews on the games they've played, share their thoughts and also indicate if they recommend the game to other users or not. Other users can indicate their thoughts about the reviews by indicating whether the review was funny, or helpful, and even give awards to users for their reviews. This information is available in the users-items-reviews dataset.

Looking at the two datasets, there is a combination of explicit and implicit indicators that we could use to calculate ratings that a user may have given to a game. We decided that we could use playtime - that is, the amount of time, in minutes, that a user has played a game for - as one of the implicit indicators to reflect the user's attitude to games. At the same time, we also extracted some useful information from the reviews that users posted and used that as another implicit indicator in the recommendation model. Finally, we extracted the information about whether a user "recommends" a game or not, to other users, and used that as an explicit indicator.

---
[1] https://newzoo.com/resources/blog/the-games-market-will-show-strong-resilence-in-2022

| Feature | Type | Description |
|---|---|---|
| User_id | str | Unique identifier of a user |
| User_url | str | URL link to a user's profile |
| Item_id | int | Unique identifier of item |
| Item_Counts | int | Number of games owned by the user |
| Item_name | str | Registered name of the item on Steam platform |
| Playtime_forever | float | The number of minutes a user has played this game since ownership |
| Playtime_2weeks | float | The number of minutes a user has played this game in the recent 2 weeks |

Table 1: Users Items Dataset

| Feature | Type | Description |
|---|---|---|
| User_id | str | Unique identifier of a user |
| User_url | str | URL link to a user's profile |
| Item_id | int | Unique identifier of item |
| Funny | int | Number of people who marked this comment as funny |
| Helpful | int | Number of people who marked this comment as helpful |
| review | str | Free text which reports the user opinion |
| posted | str | Review posted time |
| Last_edited | str | Review last edited time |
| Recommended | bool | Recommended by user (true or false) |

Table 2: Users Reviews Dataset

## 2.2 Exploratory Data Analysis

In the Users Items dataset, there are in total 5,153,209 purchasing records from a total of 70,912 different users on 10,978 different items. We can tell that the interaction matrix between users and items is very sparse, which later leads us to build the recommender system using ALS method which is good at solving issues related to scalability and sparseness of ratings data.

Since we take the playing time as one of implicit feedback metrics, we did the preliminary analysis on the playtime distribution. Out of the two metrics, we chose the '*playtime_forever*' feature as the evaluation metric to mitigate the impact of time and measure the game that manages to capture most loyal players throughout time, which could hardly be reflected in a seasonal feature like '*playtime_2weeks*'. We can tell that there are large variations when it comes to the playtime of games. There are games that are constantly played by a large player base, while at the same time an enormous number of games are rarely played or even not played at all.

Meanwhile, we calculated the average time played per user and per item respectively. We noticed that although the number of users is about seven times the number of games, the average time played per user is much lower than that of the average game played per item. It shows that the user base is exposed to a much larger base of games and thus they could choose to interact with as many games as they want. It is also worth noting that the average time played per game has a clear long-tail distribution: *"Counter-Strike: Global Offensive" (more known as CSGO)* and "*Garry's Mod*" ranked the top 2 most played games of all times at an unrivaled advantage, but such a gap is not seen for other games.

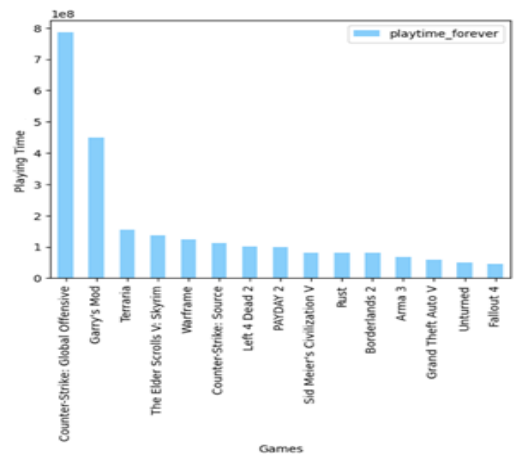

Figure 1: Total Playtime (minutes) of Most Popular Games

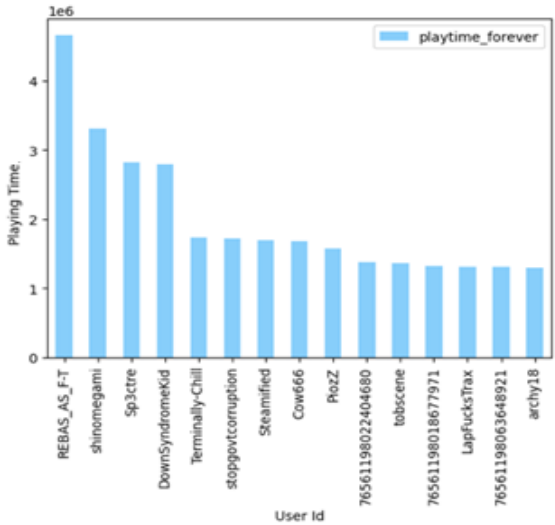

Figure 2: Total Playtime (minutes) of Most Active Users

Another important feature for the recommendation system is the actual review texts posted by users. We used Wordcloud to highlight popular words and phrases based on frequency and relevance. From the Word cloud generated, we can tell that the reviews are usually made up of game genres such as *'survival'*, user attitude like *'enjoy'* and usage scenarios like *'friend'*. Thus, it is decided that sentiment analysis will be implemented to extract users' attitude towards games as part of the input for the recommendation system.

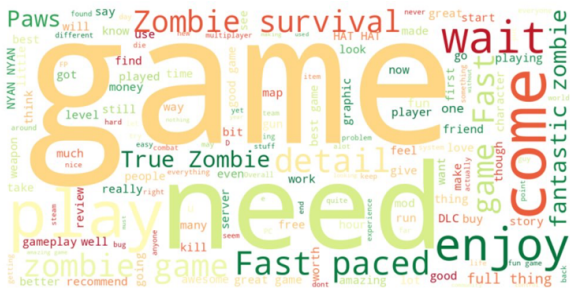

Figure 3: User Review Word cloud

## 2.3 Sentiment Analysis

With all the reviews that users wrote for different games, sentiment analysis can be performed to extract useful information on whether the user likes or hates the game. The information can later be used as part of the input for the recommendation system.

For sentiment analysis, we used the VADER SentimentIntensityAnalyzer. VADER is a lexicon and rule-based feeling analysis instrument that is explicitly sensitive to suppositions communicated in web-based media. VADER utilizes a mix of lexical highlights that are marked by their semantic directions as positive or negative. Besides, VADER not only tells about if the text is positive or negative; it also tells us concerning how positive or negative it is by using the latitude of the positive/negative score, which facilitates us in further categorizing comments into more types of sentiments.

From the pie chart below, we can tell that the majority of the reviews are positive, with around 1/5 being neutral and a very small portion of them being negative. Table 3 lists down a few examples of different sentiment categories.

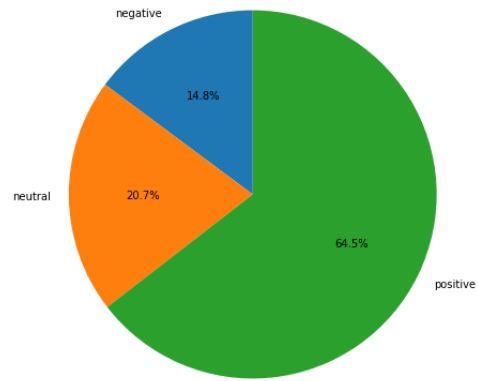

Figure 4: Pie Chart of Review Sentiment Classes

| Sentiment Class | Review | VADER Score |
|---|---|---|
| Positive | Simple yet great replayability In opinion zombie horde team work good leave 4 dead plus global leveling system Alot earth zombie splatter fun whole family Amazed sort FPS rare | 0.8402 |
| Neutral | Do buy game nothing shadow could amasing gameIf wan na se could go httpwwwbuildandshootcomserverlist_pagephpThanks jagex ing | 0 |
| Negative | RUBBISH GAME DO NOT PLAY EVEN IF IT IS FREEABSOLUTE UTTER I DINNAE KNOW WHY YOU WOULD PLAY THIS GAME WHEN ITS | -0.3964 |

| | ESSENTIALLY JUSTAN UNREAL TOURNAMENTESQUE GAME WITH ALL THE FUN SPEED REMOVED ANDARBITRARY LIMITATIONS PLACED ON VISUAL CUSTOMISATION AND WEAPONS | |

Table 3. Examples of Reviews from Each Sentiment Class:

## 2.4 Recommender System Models

A common approach to designing a recommendation system involves the concept of collaborative filtering (CF).

Collaborative filtering can be done in two ways: directly modeling the relationships between users and items, often referred to as neighborhood methods, or indirectly modeling these relationships using inferred variables (latent factors). For direct relationships, user-to-user or item-to-item based approaches can be used. In 'user-based' collaborative filtering, we intend to find a set of users most similar to the target user who has rated a particular item. In 'item-based' collaborative filtering, we intend to find a set of items most similar to the item which have been rated by a user.

Latent factors are variables that are not directly observable but assumed to have an influence on users' preferences for content. The algorithm that is used to associate user and item relationships through latent factors rather than directly representing these associations is ALS.

In the user matrix, rows represent users and columns are latent factors. In the item matrix, rows are latent factors and columns represent items. The Factorization matrix model learns to factorize the rating matrix into user and item representations, allowing the model to predict better-personalized item ratings for users.

### 2.4.1 Alternating Least Squares (ALS)

"Alternating Least Square (ALS)" is a matrix factorization algorithm that runs itself in a parallel fashion. It is implemented in Apache Spark ML and built for large-scale collaborative filtering problems. ALS is designed to resolve the issue of scalability and sparseness in ratings data, and it scales well to very large datasets while being simple. Furthermore, as described above, the ALS model stands out for being capable of working using implicit feedback. As for the input data given to the algorithm, we used columns *user_id, item_id*, and *user's ratings* about the game.

In order to derive the user ratings, we used the following three approaches:

1) Use only the 'playtime_forever' field.
2) Use the 'playtime_forever' field along with the sentiment ratings.
3) Use the 'playtime_forever' field along with the existing recommendation ratings.

To calculate the ratings about user-game interaction, we have to assume a user game interaction metric. We can assume playing time as fairly persuasive information about users' interests. We compare each individual user's playing time for a game with the game's median played time by all the users.

The exact formula for assigning the ratings between users and items based on *'playtime_forever'* field is shown in the table below:

| S.no | Criteria | Rating |
|---|---|---|
| 1. | If a user's playing time is greater than the game's median playing time from all of its users (median playing time). | 5 |
| 2. | If the user's playing time is less than median playing time but greater than 0.8 times the median playing game. | 4 |
| 3. | if the user's playing time is less than 0.8 times the median playing time but greater than 0.5 times the median playing game | 3 |
| 4. | if the user's playing time is less than 0.5 times median playing time but greater than 0.2 times the median playing game | 2 |
| 5. | if the user's playing time is less than 0.2 times the median playing time | 1 |

Table 4. Normalized rating calculation formula (using sentiments).

For the model 2, we use the sentiments (section 2.3) to normalize these ratings and then finally derive the rating to be supplied to the ALS algorithm. To elaborate further, if a user has got a rating (from the above table) of '5' for a game but the sentiment is '*negative*'. Then, the normalized rating becomes '4'. Similarly, a rating of '4' will go up to '5' in case the user leaves *'positive'* sentiment review. Furthermore, the rating will remain unchanged in case of neutral review or no review.

For the model 3, we use the *'recommended'* field present in the reviews dataset to normalize the ratings obtained (from table 4). For instance, if the playtime rating obtained is less than, or equal to 3, but the user has explicitly recommended the game, then we increase the rating by 2. However, if the rating is 4 or 5 but the recommendation from the user is 'No', then we reduce the rating by 2.

## 3 Results

Using the methodology explained in the previous part, we trained the model on the respective ratings data, each running for 10 iterations and, each time with a different number of latent factors. We summarize those results in Figures 5,6, and 7 and Table 5.

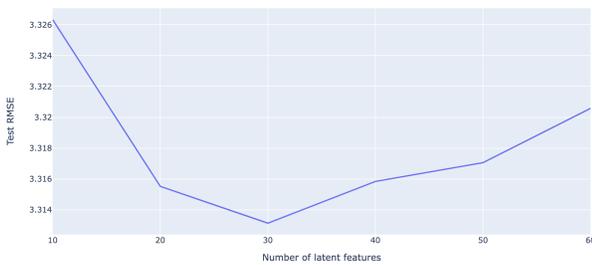

Figure 5: Approach 1- Training with different numbers of latent factors.

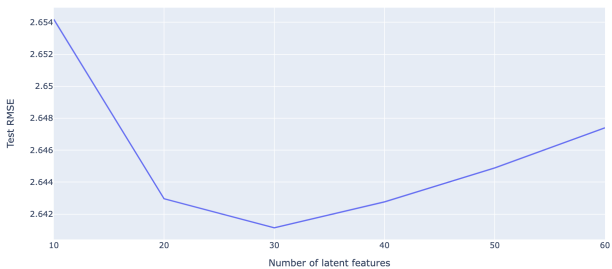

Figure 6: Approach 2- Training with different numbers of latent factors.

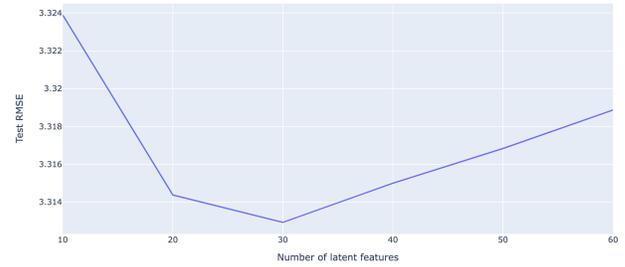

Figure 7: Approach 3- Training with different numbers of latent factors.

RMSE Value comparisons for 3 approaches:

| S.no | Approach | RMSE |
| --- | --- | --- |
| 1 | ALS (with playtime) | 3.31 |
| 2 | ALS (with playtime & sentiments) | 2.64 |
| 3 | ALS (with playtime & recommendations) | 3.32 |

Table 5: Test results for every model presented, including results with and without (WS) sentiment analysis considered.

Evaluation results show that RMSE decreases upon including the sentiments with *playtime* based ratings. However, when we include the recommendations of users in the reviews, along with the *playtime* based ratings, the RMSE doesn't change by a significant amount. This could be explained by the fact that if a user has played a game for a large amount of time, the implicit rating can go up to 5, and it'd be very unlikely that a user will not recommend a game to others if they themselves played that game for a long duration. Similarly, if a user buys a game but doesn't play it much, and also doesn't recommend it to other users, it can be explained that the user didn't like the game so didn't end up playing it much, and also doesn't recommend it to other users. There were some cases (<0.1%) where a user who didn't play a game for a large amount of hours, recommended it still and vice versa but looking at the percentage, such cases were far and few in between.

Inferring from the graph in Figure 5, we can see that the best count of latent factors in the model were found to be close to 30.

Finally, we use the best configuration for each of the model to recommend games to the users:

### User1

| S.no | Item Id | Titles |
|---|---|---|
| 1 | 253900 | Knights and Merchants |
| 2 | 253980 | Enclave |
| 3 | 7510 | X-Blades |
| 4 | 343360 | Particula |
| 5 | 214340 | Deponia |

### User2

| S.no | Item Id | Titles |
|---|---|---|
| 1 | 253900 | Knights and Merchants |
| 2 | 253980 | Enclave |
| 3 | 343360 | Particula |
| 4 | 7510 | X-Blades |
| 5 | 303390 | Dead Bits |

Table 6: Model 1 - Top 5 recommendations for 2 users.

### User1

| S.no | Item Id | Titles |
|---|---|---|
| 1 | 331710 | Why So Evil |
| 2 | 253920 | Gorky 17 |
| 3 | 343360 | Particula |
| 4 | 293180 | Overcast - W & W |
| 5 | 303390 | Dead Bits |

### User2

| S.no | Item Id | Titles |
|---|---|---|
| 1 | 331710 | Why So Evil |
| 2 | 303390 | Dead Bits |
| 3 | 293180 | Overcast - W & W |
| 4 | 343360 | Particula |
| 5 | 253920 | Gorky 17 |

Table 7: Model 2 - Top 5 recommendations for the same 2 users.

### User1

| S.no | Item ID | Titles |
|---|---|---|
| 1 | 7510 | X-Blades |
| 2 | 253900 | Knights and Merchants |
| 3 | 263980 | Out There Somewhere |
| 4 | 293180 | Overcast - W & W |
| 5 | 214340 | Deponia |

### User2

| S.no | Item ID | Titles |
|---|---|---|
| 1 | 253900 | Knights and Merchants |
| 2 | 7510 | X-Blades |
| 3 | 293180 | Overcast - W & W |
| 4 | 263980 | Out There Somewhere |
| 5 | 253980 | Enclave |

Table 8: Model 3 - Top 5 recommendations for the same 2 users.

## 4 Conclusions

In this work, we used 3 different approaches to train a recommendation system on a large dataset with an objective to recommend games to end users. We first conducted an exploratory data analysis to determine the features of key interest that could be helpful. The analysis demonstrated that *'playtime_forever'* ("the total number of minutes played on record") could be useful as an implicit feedback feature. We used it to derive ratings for different user item combinations and finally used the Alternative Least Squares (ALS) matrix factorization algorithm to recommend games to users. We further did sentiment analysis for the item reviews dataset using NLP techniques and used the corresponding sentiment scores in conjunction with the ratings derived from the *'playtime_forever'* feature to train ALS model and make recommendations. Finally, we used the data from recommendations that users explicitly made in their reviews, along with their implicit ratings to generate recommendations in the third approach.

The results show that incorporating sentiments along with other implicit factors can improve the recommendation system's performance, while using explicit recommendations made by users doesn't change the performance much because recommendations made by users are highly similar to the amount of time that a user plays a game.

## 5 Future Work

In exploratory data analysis, it turned out the user-item matrix was highly sparse with only 0.66% seen registries. Large sparsity is not ideal for training a model of interactions between users and items. Taking this fact into account, we should add in filtering before fitting the dataset into the model. In our case, filtering can be based on items such as only including games that have more than 500 times purchase records or based on users like only containing customers with at least 100 purchased items. Having a denser interaction matrix is beneficial to improve the accuracy of the model.

We can also try out other models such as Factorization Machines (FM), DeepNN and DeepFM. The latter model incorporates a DeepNN and a FM layer, which work in parallel to introduce higher-order interactions between inputs. This could further improve the accuracy of predictions by taking into account factors such as novelty, diversity, and precision.